\documentclass[epsfig,12pt]{article}
\usepackage{epsfig}
\usepackage{graphicx}
\usepackage{array}
\usepackage{color}
\usepackage{bm}
\usepackage{cite}
\usepackage{geometry}
\usepackage{latexsym}
 \usepackage{amsmath, amssymb, amscd, xypic, graphicx}

\newcommand{\beq}{\begin{equation}}   
\newcommand{\eeq}{\end{equation}}
\newcommand{\beqn}{\begin{eqnarray}}   
\newcommand{\eeqn}{\end{eqnarray}}

\begin{document}

\vspace{1cm}

\begin{center}

{  \Large \bf  Introduction to \\[2mm]
``Standing together in Troubled Times"}

\vspace{3mm}

{\large\em M. Shifman}

\vspace{4mm}

Abstract

\end{center}

\begin{quote}

{\small This Introduction opens the book {\sl Standing Together in Troubled Times} which presents
a story of friendship between Wolfgang Pauli, one of the greatest theoretical physicists of the 20th century, and Charlotte Houtermans. They met at the very onset of the quantum era, in the late 1920s in Germany where Charlotte was a physics student at G\"ottingen University. At that time G\"ottingen was right at the heart of groundbreaking developments in physics. Both Pauli and Houtermans personally knew major participants in the quantum revolution.

Caught between two evils Ê-- German National Socialism 
and Soviet Communism -- Charlotte Houtermans would have likely perished if it were not for the brotherhood of physicists: Niels Bohr, Wolfgang Pauli, Albert Einstein, James Franck, Max Born, Robert Oppenheimer and many other noted scientists who tried to save friends and colleagues (either leftist sympathizers or Jews) who were in mortal danger of becoming entrapped in a simmering pre-WWII Europe.

This book is based on newly discovered documents from the Houtermans family archive, including Pauli's numerous letters to Houtermans, her correspondence with other great physicists, her diaries, and interviews with her children. Almost all documents presented in this book are being published for the first time.}
\end{quote}

This book grew from my work on the collection {\sl Physics in a Mad World} \cite{PMW} which was largely devoted to physics and physicists who had the misfortune to live and work under the Nazi regime in Germany and the communist regime in the USSR. I used to think that I knew all about the atrocities committed by these regimes from history textbooks and abundant literature. It turned out that tracing the destinies of physicists 
whose theories I explain to my students every year created a much more personal picture, adding a totally new -- human -- dimension to the tragic events triggered by two of the most disastrous experiments in social engineering that shaped the history of the 20th century. 

The 2015 Nobel Prize laureate in literature, Svetlana Alexievich, entitled  her book 
{\sl War's Unwomanly Face}. She characterizes it as the testimony of women recollecting their past, 
on how girls who dreamed of becoming brides, became soldiers in 1941. They had to kill  the enemy who had attacked their homes and homeland with unprecedented cruelty.  

Reading Alexievich's book I thought that, perhaps, its title was not quite accurate. Through 
the ugly faces of war I saw the human faces of people -- young men and women -- who were sent to
slaughter against their will, by their ruthless dictators, with no regard for human lives. Dictatorships may pursue different ideologies  but under a closer examination  they are all based on the presumption 
 that the end justifies the means, no matter how horrific the means might be...  Still,
despite all this,  friendship, love and compassion survive even under these inhuman circumstances.

This is the nature of the book you are now opening. It is about the friendship between  Wolfgang Pauli, one of the greatest physicists of the 20th century,  and Charlotte Houtermans.
They met at the very onset of the quantum era, in the late 1920s in Germany where Charlotte was a physics student at G\"ottingen University. At that time G\"ottingen was right at the heart of groundbreaking developments in physics. Both Pauli and Houtermans personally knew major participants in the quantum revolution.
In the 1920s and '30s the emergent quantum world was very much central Europe: Germany, Austria, Hungary, Denmark, and Switzerland. 

Both Wolfgang Pauli  and Charlotte Houtermans went through trials and tribulations so abundant at the time of the clash of the two barbarian ideologies and two dictatorships which served them. Newly found letters from Pauli, 
Einstein,\index{Einstein, Albert} Franck,\index{Franck, James} Oppenheimer\index{Oppenheimer, Robert} and others to Charlotte give a valuable and rare insight  into physicists' relationships beyond science, in troubled times.

Wolfgang Pauli was a great physicist, a trailblazer of the quantum era whose life is well documented. 
Pauli's truculent style of scientific discourse gave rise to legends.
However, some aspects of Pauli's human side are less known to the general public. His  letters to Charlotte Houtermans, their life-long friendship,  show in more than one way that Wolfgang Pauli was a man of warm heart, a tender and caring friend who tried to help his friends whenever they needed help and whenever he could. This is a precious addition to Pauli's scientific biography (see \cite{pscientificbio}) revealing to us Pauli the human being.

In a broader context this book is about a brotherhood of physicists. Charlotte Houtermans who found herself  between two evils  -- Soviet communism and German National Socialism -- would have probably perished if it were not for this brotherhood. It was not a deliberately organized society, nor a formal organization. Rather, professional physicists and people related to physics all over the world acted on impulse, out of the kindness of their hearts,
in an attempt to save or help their colleagues who found themselves entrapped in simmering pre-war Europe. 

Charlotte's husband Friedrich (Fritz) Houtermans\index{Houtermans, Fritz}, a  German
physicist who suggested that the source of stars'
energy was thermonuclear fusion, in early 1935
fled to the Soviet Union in an
attempt to save his life from Hitler's Gestapo. Fritz Houtermans  who had been a member of the
German Communist Party since 1926 could expect no mercy from the Nazis.
Charlotte followed him with their daughter Giovanna born in Berlin in 1932. Half a century later 
Annika Fjelstad,\index{Fjelstad, Annika} Charlotte's granddaughter, wrote: ``two dreamers
in Berlin, the political womb of an unborn war."

Fritz Houtermans\index{Houtermans, Fritz}, took an
appointment at the Ukrainian Physico-Technical Institute in Kharkov 
and worked there for three years. In the Great Purge of 1937, he 
was arrested by the NKVD (the Soviet Secret Police, the KGB's
predecessor) in December 1937. He was tortured and confessed to being a Trotskyist
plotter and German spy, out of fear of threats against his wife
 and children (his son Jan was born in Kharkov, USSR, in 1935).
 
The story of Charlotte's escape from the USSR, with her two children (see Chapter 4), belongs in a Hollywood thriller. In the last days of 1937 she managed to escape from Moscow to Riga, Latvia. However, this was only the beginning of her long journey out of turbulent  Europe to a new life in the New World. Niels Bohr helped her to reach Copenhagen, Denmark. Many physicists -- from Bohr's\index{Bohr, Niels} colleagues in Copenhagen to Patrick 
Blackett\index{Blackett, Patrick} in England -- were instrumental in Charlotte's relocation to London. 
While she could not find any job in England, her friends Bohr, Blackett and especially Robert 
Oppenheimer,\index{Oppenheimer, Robert}
as well as the Academic Assistance Council, supported her financially. The most outstanding physicists, such as 
Fr\'ed\'eric Joliot,\index{Joliot, Fr\'ed\'eric} Albert Einstein,\index{Einstein, Albert} James Franck,\index{Franck, James} Max Born,\index{Born, Max}
and many others joined the fight for Fritz Houtermans' release. Alas... to no avail. Her friends helped her emigrate to the United States where she worked as a physics educator for the rest of her life.  There she appealed to the First Lady, Mrs. Eleanor Roosevelt,\index{Roosevelt, Eleanor}  who became interested in the fate of Charlotte's husband and contacted 
Soviet authorities at various levels. Charlotte's correspondence with Mrs. Roosevelt
is also published in this book.

In the aftermath of the Molotov-Ribbentrop pact the NKVD turned Fritz Houtermans
over to... the Gestapo in Nazi Germany (in May 1940). Thus, he traded Soviet prisons for a prison in Berlin.

 In the United States mortal danger  for Charlotte and her children was over, but not her problems: 
 starting a new life from scratch, establishing herself as a college professor, Giovanna's and Jan's cultural adaptation, 
 a painful and highly unjust divorce from Fritz -- to name just a few...   
 She overcame all these obstacles with the help of her friends. Wolfgang Pauli was among them. 

 \begin{center}
 ***
 \end{center}
 
Wolfgang Pauli, the 1945 Nobel Prize winner in physics,  was known among his 
colleagues not to be an easy or forthcoming person to deal with. He applied extremely high criteria of ``cleanliness" both to his own works and to those of other theoretical physicists and was not afraid of open conflicts in those cases when he saw gaps or imperfections in the line of reasoning. 
Peter Freund once called him \cite{pf}  ``an inquisitor defending physics," 

\begin{quote}

Now mediator bringing his friends to their senses, now
merciless critic of the hopeless dead end, now spoilsport
who discouraged many a major discoverer, now brilliant
discoverer himself, Wolfgang Pauli hovers over his contemporaries as a kind of 
conservative and thoroughly honest
supreme judge, an inquisitor defending physics. His colleagues dubbed him ``the conscience of physics." 
\end{quote}

\index{Freund, Peter}
 Abraham Pais \index{Pais, Abraham} recollects \cite{pais}:

\begin{quote}

[In 1946 Pauli] had already long been recognized
as one of the major figures in twentieth-century physics, not only because of his
own contributions, but also because of his critical judgments -- which could be
quite sharp, but nearly always to the point -- of others' work. He was known as
the conscience of twentieth-century physics, as is reflected in his voluminous
correspondence, a very rich source of information concerning the development
of physics in the first half of the twentieth century [...] 
His letters are nearly all in German, which he wrote masterfully.

\end{quote}

This feature of Pauli's character as a physicist -- his sharply critical attitude to his colleagues' work -- is documented in many scientific biographies, see  \cite{pscientificbio}. 
This might have created an impression of a negative aura around him. Moreover, sometimes Pauli's biographers place an emphasis on a remark of his that ``womenÉ [are] pleasant things to play with, but not something to take seriously." I hope that Pauli's letters to Charlotte published in this book
will persuade the reader that  the above remark was taken out of context.  
Pauli certainly had a complicated personality. His character was not ``one-dimensional." Rather it was a combination of a deep love of physics and determination to defend it, with sincerity, humanism, and  kindness.   

Pauli's critical mind could not bear unsound results, incomplete works, or hand-waving arguments.
It was important that he applied the same high criteria to his own results. Very instructive in this respect is the story of his last work with Heisenberg which remained unpublished. 

After Germany's defeat in WWII, Heisenberg\index{Heisenberg, Werner} found himself in scientific isolation, especially among American colleagues. If before the war the front-line of scientific research in physics
belonged to Germany, after the war the physicists working in the US defined the cutting edge. 
 I think that  Heisenberg made a deliberate choice to distance
himself from ``popular" topics on which the major efforts of the American theoretical community
were focused. Naturally, such topics were also points of attraction for young and ambitious researchers.

Instead, in the mid-1950s Heisenberg embarked on the investigation of a ``unified field theory of elementary particles" based on four-fermion interactions. Heisenberg's idea was to use a fundamental Dirac field, endowed with a four-fermion
interaction of a special type, to write and  solve the emerging nonlinear
field equations. He hoped to get in this way a complete set of ``elementary" particles known
at that time (both, hadrons and leptons) and dynamically describe their properties in terms of
one or two input constants. From today's perspective it is absolutely obvious that this direction 
was a dead\index{Volkov, Dmitry} end.\footnote{In 1966, Heisenberg published a book \cite{pHeiBook} where he summarized his work on ``unified field theory."
Although Heisenberg himself did not want to admit that his attempt was a failure, others did; 
 I do not think that many people in the world studied this book carefully. In fact, it
was to a large extent obsolete by the time of its publication.
One might say that the whole program was a wasted effort. Well, perhaps not all of it...

On the other side of the Iron Curtain, Heisenberg's book ignited the imagination of Dmitry
Volkov \cite{pya}, a theoretical physicist from Kharkov (Ukraine) who used it in an indirect way, as an impetus to developing nonlinear supersymmetry
and then supergravity, see \cite{pVolkov}.}
Not only could it  not be made viable theoretically, but it also contradicted
experimental data which started appearing in the 1960Õs. The man who 
carried out this research  does not seem to be the Heisenberg of the pre-WWII time.

In late fall of 1957 Heisenberg came to see Pauli in Zurich in search of Pauli's 
mathematical advice on one of the aspects of his (Heisenberg's) ``unified field theory." 
For over a year Pauli had resisted 
Heisenberg's previous invitations to collaborate on this topic. But this time Heisenberg was more insistent. 
For reasons unclear  to me, Pauli got involved in Heisenberg's construction.  As a result, Heisenberg
prepared a joint preliminary report which, although unpublished, is reprinted in Heisenberg's {\sl Collected Works} \cite{PauHei}. On January 20, 1958, Pauli wrote \cite{pscientificbio}: ``The preliminary report  which 
G\"ottingen now sends out should not yet be printed in this form. Surely it still contains mistakes in the detail."  On February 1 and 2, 1958, Pauli wrote to Heisenberg from New York, mentioning the discussions he had  after his  seminar at Columbia University, and insisting on postponing
their joint publication. In a month, Pauli came to the conclusion that
he could no longer participate in the dubious endeavor initiated by Heisenberg. 
On April 7 Pauli announced to Heisenberg his final decision: 

\begin{quote}

I must totally  drop the plan to publish with you a work ``On the Isospin Group in the Theory of
Elementary Particles."
\end{quote}

On April 8, 1958, he sent the following circular letter to all his 

\vspace{1mm}

\noindent
colleagues who had received the preprint:
\begin{quote}

As essential parts of the preprint with the above title  don't agree any longer with my opinion, I am forced to give up the plan to publish a common paper with Heisenberg on the subject in question. Particularly, I am now convinced that the degeneration of the vacuum should not be used in order to explain the possibility of a half-integer difference between ordinary spin and isospin for some strange particles. The idea of a unification of the spinor field seems to fail here and I believe that one should try to introduce, besides spinors with isospin 1/2 either other spinors with isospin 0, or at least one scalar field with isospin 1/2 (``Goldhaber\,\footnote{Maurice Goldhaber (1911-2011) was an American physicist of Austrian-Jewish descent. In the 1950s M. Goldhaber proposed (with Edward Teller) that the so-called ``giant-dipole nuclear resonance" was due to the neutrons in a nucleus collectively vibrating against the proton component.}
\index{Goldhaber, Maurice} model"), in order to reach an interpretation of the elementary particles.
\end{quote}

 \begin{center}
 ***
 \end{center}

In working on this book, I read some of Pauli's and Heisenberg's\index{Heisenberg, Werner} works, 
original publications, and review articles of other authors released in the 1950s. Since that time, quantum field theory underwent two profound revolutions
which completely changed its face: (i) the discovery of Yang-Mills\index{Yang, Chen Ning}\index{Mills,  Robert} theories \cite{pYang:1954ek} and their asymptotic freedom \cite{pGWP}, and (ii) the discovery of supersymmetric field theories \cite{psusy}. These discoveries happened in 1954 and the 1970s, respectively. 
Surprisingly, Wolfgang Pauli could have been a pioneer in both revolutions were it not for his supercritical attitude to incomplete works. In fact, he ``discovered" Yang-Mills theories before Yang and Mills but did not publish because at the time of discovery he did not know what to do with massless vector fields (Yang and Mills just ignored this question). Moreover, as early as in 1950s,  Wolfgang Pauli delivered a landmark series of lectures at the Swiss Federal Institute of Technology (ETH) in Zurich. They were published in English by MIT Press only in 1973.
In Section 9 of Volume 6 \cite{PauliLec} Pauli discusses the vacuum energy density in various field theories
known at that time. He observes that adding the Dirac spinor contribution to that of two complex scalar fields cancels divergences and produces zero vacuum energy -- the first ever hint to supersymmetry! 

 \begin{center}
 ***
 \end{center}

This volume consists of four parts and six chapters. Part I is devoted to Wolfgang Pauli and Charlotte Houtermans,
the two main characters in the narrative which follows in the main body of the book.
 
Of course, every physics student knows that Pauli was a great physicist who invented the exclusion principle and predicted the existence of neutrinos. However, details of his personal biography are known to a lesser extent. In Chapter 1, I briefly 
summarize Pauli's life, both scientific and non-scientific, with the emphasis on the latter. His life journey was by far not as smooth as it might seem to young people now. 

The story of  Charlotte Houtermans who at the crucial moments of her life  found herself at the epicenters of quantum revolution in physics (in G\"ottingen, Berlin and Copenhagen), and the social cataclysms in Europe, was practically unknown to the western reader until recently. The first publications in English appeared a few years ago \cite{PMW,pAmaldi}. Chapter 2 narrates Charlotte's biography which I have compiled using various sources: Charlotte's diaries and other documents from her personal archive, recollections of her children, Giovanna Fjelstad\index{Fjelstad, Giovanna}  and Jan Houtermans,\index{Houtermans, Jan} archival documents from Russia and elsewhere, and, finally,  memoirs of people who knew her. 

Part II (Chapter 3) presents a collection of Pauli's letters to Charlotte dating from December 31, 1937, to 
February 16, 1942.  In Chapter 3, I also included a letter from James Franck\index{Franck, James} to Pauli of October 31, 1937, and the first half of Pauli's letter to Weisskopf of January 13, 1938. The latter had been published in full previously. I thought, however, that presenting a fragment of this letter in English would 
give the reader a more complete idea of the events in my narrative. 

Part III presents Charlotte's {\em Recollections} (Chapter 4). To be more exact, it is one of a few manuscripts she
prepared in the 1960s and later. Although in some parts they overlap with each other they are not identical. 
A different version of Charlotte's diary was published in \cite{PMW}.

Charlotte Houtermans was a talented writer. It is a pity she never published her memoirs and stories when she was alive. Well... as they say, better late than never.

In Chapter 5, additional (i.e. other than Pauli's)  previously unpublished letters to and from Charlotte are collected. Here the reader will find letters from Albert Einstein,\index{Einstein, Albert} Patrick Blackett,\index{Blackett, Patrick}
Max Born,\index{Born, Max} James Franck,\index{Franck, James} Max von Laue,\index{Laue, von, Max} Robert Oppenheimer,
\index{Oppenheimer, Robert} Christian M{\o}ller\index{M{\o}ller, Christian} (Bohr's assistant), Eleanor Roosevelt,\index{Roosevelt, Eleanor}  and others.
To my mind, they are of broad interest not only to historians of science but to the general public as well.

Finally, in Part IV (Chapter 6), the German originals of some  Pauli's letters are reproduced.

Footnotes in this book are mine if not stated to the contrary.

\vspace{10mm}

 \begin{center}
{\large\bf Acknowledgments}
 \end{center}
 
First and foremost, I am deeply grateful to Giovanna Fjelstad and Jan Houtermans, the children 
of  Charlotte Houtermans, who made available 
to me her personal archive, and kindly gave their permission to publish relevant letters and photographs.
Their remarks and suggestions were very important.

I am indebted to Prof. Gerhard Ecker,\index{Ecker, Gerhard} who translated from German Pauli's letters  (and, in addition, two of Einstein's and two of Franck's letters) and acted as my adviser. The reader will find his comments in the book.  Other letters were translated from German by Alexander Tschernow\index{Tschernow, Alexander},
Annika Fjelstad, Ilze Mueller, and Roman Zwicky to whom I would like to say thank you. 

I acknowledge many pleasant and fruitful conversations with Giovanna  and Jan, and Giovanna's daughter
Annika Fjelstad.\index{Fjelstad, Annika} I would like to thank Jean Richards, Irena 
Gross, Karen Kettering,  John Schlesinger, Alexander Khodjamirian, Michlean Lowy Amir, Michael Vinegrad,\index{Vinegrad, Michael}
Micha\l{} Prasza\l{}owicz,  Marina Ilyushina, 
Tasya Tschetschik,  and  Roman Zwicky  for useful communications. Annika Fjelstad,
RaeAnna Buchholz, Adam Peterson, Jean Richards, Jane Gatrell, and Gina Ristani were my invaluable helpers as far as English grammar is concerned.

Graphic design was done by Polina Tylevich\index{Tylevich, Polina} who had been also responsible for the design of some previous books of mine (e.g. \cite{pLandau}). I appreciate her contribution. Thank you, Polina. 

As usual, I would like to thank my World Scientific contact, Lakshmi Narayanan, for her generous assistance.


\end{document}